\documentclass[11pt]{article}
\usepackage{amsmath,amssymb, color,cite}

\usepackage{mathrsfs}
\usepackage{slashed}

\usepackage{amsfonts}
\usepackage{cleveref}
\DeclareMathOperator{\Tr}{Tr}

\makeatletter
\@addtoreset{equation}{section}
\makeatother

\newcommand{\be}{\begin{equation}}
\newcommand{\ee}{\end{equation}}
\newcommand{\bea}{\setlength\arraycolsep{2pt} \begin{eqnarray}}
\newcommand{\eea}{\end{eqnarray}}

\def\ndelta{\delta\hspace{-0.50em}\slash\hspace{-0.05em} }

\def\0{{\sst{(0)}}}
\def\1{{\sst{(1)}}}
\def\2{{\sst{(2)}}}
\def\3{{\sst{(3)}}}
\def\4{{\sst{(4)}}}
\def\5{{\sst{(5)}}}
\def\6{{\sst{(6)}}}
\def\7{{\sst{(7)}}}
\def\8{{\sst{(8)}}}
\def\sst#1{{\scriptscriptstyle #1}}

\allowdisplaybreaks

\begin{document}

\begin{flushright}
\hfill{}

\end{flushright}

\vspace{25pt}
\begin{center}
{\Large {\bf Linear Newman-Penrose charges as subleading BMS and dual BMS charges}}

\vspace{35pt}
{\bf George Macaulay }

\vspace{15pt}

{\it School of Mathematical Sciences,
Queen Mary University of London, \\
Mile End Road, E1 4NS, United Kingdom.}

 \vspace{35pt}

\date\today

\vspace{20pt}

\underline{ABSTRACT}
\end{center}

\noindent  In this paper, we further develop previous work on asymptotically flat spacetimes and extend subleading BMS and dual BMS charges in a large $r$ expansion to all orders in $r^{-1}$. This forms a complete account of this prescription in relation to the previously discovered Newman-Penrose charges. We provide an explanation for the origin of the infinite tower of linear Newman-Penrose charges with regards to asymptotic symmetries and justify why these charges fail to be conserved at the non-linear level as well as failing to exhibit full supertranslation invariance even at the linear level.

\noindent

\thispagestyle{empty}

\vfill
E-mail: g.long@qmul.ac.uk

\pagebreak

\section{Introduction} \label{Intro}
In 1968, Newman and Penrose considered a tower of charges that were shown to be conserved in linearised gravity for asymptotically flat spacetimes obeying particular fall-off conditions \cite{NP}. In the full non-linear theory, the tower collapses and only the first 10 charges are conserved. Surprisingly, these charges are conserved even in the presence of flux at null infinity \cite{bondi,sachs,WZ}. It is expected that no further such quantities can be constructed that are non-linearly conserved without imposing restrictions on the metric i.e.\ removing flux at null infinity. It was later shown in Refs.\ \cite{kroon1998conserved,valiente1999logarithmic} that these charges still exist for the broader class of polyhomogeneous spacetimes of \cite{chrusciel}, with significantly weaker fall-off conditions. It is hence clear that these 10 charges are a prominent feature of asymptotically flat spacetimes with their significance being demonstrated in relation to Aretakis charges on extremal horizons \cite{aretakis2012horizon,bizon2013remark,lucietti2013horizon,us}. 

Recently, the Hamiltonian origin of these charges has been derived \cite{fakenews,dual0,dualex,Godazgar:2020gqd,Godazgar:2020kqd,godazgar2022higher} in the covariant phase space formalism \cite{lee1990local,iyer1994some,WZ} (see also \cite{durka2011gravity,durka2012immirzi,corichi2014hamiltonian,jacobson2015black,frodden2018surface}). The 10 non-linear charges can be understood as subleading BMS and dual BMS charges for Bondi backgrounds \cite{fakenews,dualex} (see also \cite{conde}), as well as for polyhomogeneous spacetimes \cite{godazgar2020bms}.  Furthermore, the investigation of \cite{Godazgar:2020gqd} also demonstrated that contributions to the Einstein-Hilbert action in the tetrad formalism of general relativity that do not affect the equations of motion may still contain important physics and should not be neglected. In particular, the Holst term \cite{nieh1982identity,holst1996barbero} is trivial by the equations of motion, but gives rise to half of the Newman-Penrose charges as well as the leading order dual charges containing information about the NUT parameter \cite{taub1951empty,newman1963empty} and therefore the topology of the spacetime (see for example the recent work of \cite{ciambelli2021topological}).

The tetrad formalism not only explains the origin of these charges in the context of the asymptotic symmetry group for asymptotically flat spacetimes - the BMS group - but can also be adapted to demonstrate that the charges obey the stronger property of supertranslation invariance, referred to as \textit{absolute conservation} by Newman and Penrose.

The infinite tower of linear Newman-Penrose charges continues to be a topic of interest (see for example \cite{freidel2021higher}), but is yet to have been investigated in this formalism. This is the purpose of this paper. In Section \ref{prelim}, we begin by reviewing the gauge choices one can make in order to study asymptotically flat spacetimes. We will also choose specific fall-off conditions for the metric functions consistent with those of Newman and Penrose. In Section \ref{SecEinsteinEqn}, we obtain the Einstein equations in our gauge and review their structure with regard to initial data.  We then obtain the linearised equations at each order in the radial coordinate $r^{-1}$. In Section \ref{Charges}, we review the BMS group and summarise the charge variations obtained in the first order formalism arising from the Palatini and Holst terms. In Section \ref{LinearNP}, we obtain an expression for the linear Newman-Penrose charges in our gauge choice and demonstrate conservation with respect to $u$, but the absence of supertranslation invariance. Finally, in Section \ref{NLNP}, we demonstrate that the non-linear Newman-Penrose charges, beyond the first 10, are not conserved at any order. We conclude with a brief discussion. 
\\

\noindent \textbf{Notation:} Latin indices $(a, b,...)$ denote internal Lorentz indices $0,1,2,3$ and are raised and lowered with respect to the Lorentz metric $\eta_{ab}$. Indices $i,j,...$ refer to the $2,3$ components specifically. We use Greek letters $(\mu, \nu,...)$ to denote the spacetime indices.  The vierbein $e^a= {e^a}_{\mu} dx^{\mu}$. The spacetime metric can be expressed in terms of the vierbein as ${ds^2=g_{\mu \nu} dx^\mu dx^\nu= \eta_{ab} e^a e^b}$.  Indices $I,J,...$ will denote spacetime indices on the round 2-sphere and will be lowered and raised using the round 2-sphere metric $\gamma_{IJ}$ and its inverse, respectively, except where explicitly stated otherwise.

\section{Preliminaries} \label{prelim}

\subsection{Gauge choices}
Let $(u,r,x^I=\{\theta,\phi\})$ be coordinates on our spacetime manifold such that the metric takes the form \cite{bondi,sachs}
\begin{equation} \label{metric}
ds^2 = -F e^{2\beta} du^2 -2e^{2\beta} du dr +r^2 h_{IJ} (dx^I-C^I du)(dx^J-C^J du),
\end{equation}
where $r$ is a radial coordinate. The spacetimes we are considering here are asymptotically flat and we use the Bondi definition of asymptotic flatness where the metric functions obey the following fall-off conditions
\begin{align} \label{metricexpansions}
F(u,r,x^I) &= 1+\sum^{N-3}_{n=0} \frac{F_n(u,x^I)}{r^{n+1}}+o(r^{-(N-2)}), \notag\\
\beta(u,r,x^I) &= \sum^{N-4}_{n=0} \frac{\beta_n(u,x^I)}{r^{n+2}}+o(r^{-(N-2)}),  \notag\\
C^I(u,r,x^I) &= \sum^{N-3}_{n=0}\frac{C_n^I(u,x^I)}{r^{n+2}}+o(r^{-(N-1)}), \notag\\
h_{IJ}(u,r,x^I) &= \sum^{N-2}_{n=0} \frac{\bar{H}_{nIJ}(u,x^I)}{r^n}+o(r^{-(N-2)}),
\end{align}
where $\bar{H}_{0 IJ}=\gamma_{IJ}$, the round metric on the 2-sphere. Here, we have assumed an analytic expansion of the metric up to a certain order.\footnote{$N$ is closely related to the \textit{degree of smoothness} used by Newman and Penrose.} We define $H_{nIJ}\equiv \bar{H}_{n\langle IJ\rangle}$ where $\langle, \rangle$ on a pair of indices denotes the symmetric trace-free part. Hence
\begin{equation}
\bar{H}_{nIJ}=H_{nIJ}+\frac{1}{2} (\Tr \bar{H}_n) \gamma_{IJ}.
\end{equation}
We can use residual gauge freedom to set $\det h = \det \gamma$. The Cayley-Hamilton theorem implies that this is equivalent to requiring
\begin{align} \label{CH}
h_{IK}h^K_J- (\Tr h) h_{IJ} + \gamma_{IJ} &=0,
\end{align}
where indices are raised and lowered with  $\gamma$. Plugging the expansion \eqref{metricexpansions} into \eqref{CH}, contracting with $\gamma^{IJ}$ and considering the coefficient of $r^{-n}$, one deduces that $\Tr \bar{H}_n$ can be expressed in terms of $\bar{H}_i$ with $i=1,...,n-1$
\begin{align}
\Tr \bar{H}_n &= \frac{1}{2} \sum^{n-1}_{i=1} \bigg{(} \bar{H}_{iIJ}\bar{H}_{n-i}^{IJ}-(\Tr\bar{H}_i) (\Tr\bar{H}_{n-i})\bigg{)}.
\end{align}
Hence, the two degrees of freedom of $h_{IJ}$ after gauge-fixing, are parameterised at each order by $H_{n IJ}$, the traceless part. Furthermore, this is the truly linear part of $h_{IJ}$ at each order.
\\
Contracting \eqref{CH} with the inverse $h^{-1}$, one deduces that
\begin{align}
h^{-1}_{IJ} &= -h_{IJ} +(\Tr h )\gamma_{IJ}
\end{align}
and so, in particular,
\begin{equation}
h^{-1}_{IJ} \big{|}_{r^{-n}} = -H_{nIJ} +\frac{1}{2} (Tr \bar{H}_n) \gamma_{IJ},
\end{equation}
which will be useful for our calculations.

Following Newman and Penrose, we will assume the Newman-Penrose scalar
\begin{equation}
\Psi_0 = \mathcal{O}(r^{-5}),
\end{equation}
which implies that $H_{2IJ}=0$. This ensures that the Newman-Penrose scalars obey the peeling property  \cite{NP61}.

\subsection{Null frame}
In order to calculate the Einstein equations in a form where they are covariant on the 2-sphere, we require a Lorentz gauge choice. Let our frame fields be
\begin{equation} \label{vierbein}
e^0 = \tfrac{1}{2} F du +dr, \quad e^1 = e^{2\beta} du \quad \text{and} \quad e^i = r E^i_I (dx^I - C^I du), \\
\end{equation}
where $E^i_I$ is a zweibein associated with the 2-metric $h_{IJ}$, so that $(h^{-1})^{IJ} E^i_I E^j_J= \eta^{ij}$ and $\eta_{ij} E^i_I E^j_J = h_{IJ}$. The inverse of the frame fields are
\begin{equation} \label{vierbein2}
e_0 = \partial_r, \quad e_1 = e^{-2\beta} (\partial_u -\tfrac{1}{2}F \partial_r +C^I \partial_I) \quad \text{and} \quad e_i =\tfrac{1}{r}E_{iJ}(h^{-1})^{IJ} \partial_I. \\
\end{equation}
In contrast to all other fields, $I,J,...$ indices on $E^i_I$ will be raised and lowered with $h_{IJ}$. For more details, we refer the reader to Refs.\ \cite{fakenews, Godazgar:2020kqd}. As in the Newman-Penrose formalism, the components of the Einstein tensor can then be written as follows
\begin{align} \label{EinsteinNP}
G_{ab} = e_a^{\mu} e_b^{\nu} G_{\mu\nu}.
\end{align}
There are different choices for $E^i_I$ and we will follow Ref.\  \cite{godazgar2022higher} and pick $E^i_I={X_I}^J \hat{E}^i_J$ with $X_{IJ}(u,r,x^I)$ a symmetric tensor on the 2-sphere. One can express $X_{IJ}$ as an $r^{-1}$ expansion in terms of $H_{nIJ}$ which means that all Einstein equations are covariant on the 2-sphere. This choice does not affect our final results.

\section{Einstein equations} \label{SecEinsteinEqn}

\subsection{Structure of the Einstein equations and initial data} \label{EinsteinStructure}

In this section, we will assume the fall-off conditions of the metric functions
\begin{gather}
F=1+o(r^{0}), \quad \beta = o(r^{-1}), \quad C^I = o(r^{-1}) \quad \text{and} \quad h_{IJ}=\gamma_{IJ} + o(r^0),
\end{gather}
but we will generally not need to assume the subleading analytic expansion for the functions in \eqref{metricexpansions}. In particular, the results below will hold for the polyhomogeneous case considered in Ref.\ \cite{godazgar2020bms}. Throughout, we will assume the following decay of the energy-momentum tensor
\begin{gather} \label{EinsteinFallOff}
 T_{00} = o(r^{-N}), \quad T_{01} = o(r^{-N}), \quad T_{0i} = o(r^{-N}), \quad T_{ij} = o(r^{-(N-1)}), \notag\\
 T_{11} = o(r^{-2}) \quad \text{and} \quad T_{1i} = o(r^{-3}),
\end{gather}
for a fixed $N\geq 6$. As discussed in for example Ref.\ \cite{fakenews}, one cannot assume the components of $T_{ab}$ obey independent fall-off conditions, so in reality, the fall off of $T_{11}$, for example, will typically be much faster. However, any further equations obtained here do not provide new information, so we do not consider them.

As discussed in Ref.\ \cite{godazgar2019taub}, the vacuum Einstein equations, in this form, fall into three categories. With the spacetime foliated by $u= const$ hypersurfaces, the $G_{00}, G_{01}$ and $G_{0i}$ equations contain no $u$ derivatives and can be viewed as first order differential equations in $r$ for $\beta, F$ and $C^I$, making them \textit{hypersurface equations}. The $G_{ij}$ equations are first order differential equations in $u$ and determine how $h_{IJ}$ evolves and are hence viewed as \textit{evolution equations}. For \eqref{metricexpansions}, this determines the evolution of $H_{(n\geq 3) IJ}$, with $H_{1IJ}$ unconstrained, constituting free data. The $G_{11}$ and $G_{1i}$ equations are also first order differential equations in $u$, but are satisfied on $r=const$ hypersurfaces, meaning that if they are true on one such hypersurface, they hold everywhere. These are therefore viewed as \textit{conservation equations}. For the metric expansions in \eqref{metricexpansions}, $F_0$ and $C_1^I$ cannot be determined from the hypersurface equations, but one can determine their evolution through the conservation equations. To summarise, one prescribes initial data $\{F_0(u_0,x^I), C_1^I(u_0,x^I), H_{(n\geq 3)IJ}(u_0,x^I)\}$ and an arbitrary trace-free tensor $H_{1IJ}(u,x^I)$
and uses the hypersurface equations to determine all quantities in \eqref{metricexpansions} on the hypersurface $u=u_0$. Next, the evolution and conservation equations are integrated to find $\{F_0, C_1^I, H_{(n\geq 3)IJ}\}$ at the next time step, with the process then being iterated.

We begin by finding the hypersurface equations. Consider first the $00$ component of the Einstein equations. We find that
\begin{equation}
G_{00}= \frac{4}{r} \partial_r \beta +\frac{1}{4} \partial_r h_{IJ} \partial_r (h^{-1})^{IJ}.
\end{equation}
Assuming $G_{00}=o(r^{-N})$, we then deduce
\begin{equation}
\partial_r \beta = -\frac{1}{16}r \partial_r h_{IJ} \partial_r (h^{-1})^{IJ}+o(r^{-(N-1)}), \label{beta}
\end{equation}
so $\beta$ can be expressed entirely in terms of $h_{IJ}$. This is a first order differential equation for $\beta$ in $r$ with the integration constant changing $\beta$ up to an $\mathcal{O}(r^0)$ term, but this contribution is zero by the fall-off condition $\beta= o(r^{-1})$, so $\beta$ is completely determined in terms of $h_{IJ}$.

Next, we compute
\begin{align}  \label{C}
G_{0i}& = E^I_i \bigg{(} \frac{1}{2r^3} \partial_r \big{(}r^4 e^{-2\beta} h_{IJ} \partial_r C^J\big{)} -r\partial_r\big{(}\frac{1}{r^2} D_I \beta\big{)}\notag \\
 &\hspace{40mm} +\frac{1}{2r}D_J \big{(}(h^{-1})^{ JK} \partial_r h_{IK} \big{)} +\frac{1}{4r} D_I h_{JK} \partial_r (h^{-1})^{JK} \bigg{)},
\end{align}
where $D_I$ is the covariant derivative associated with the round 2-sphere metric $\gamma_{IJ}$. Assuming $G_{0i}=o(r^{-N})$ and the Einstein equation for $\beta$, we can, in principle, write $C^I$ as an expression involving only $h_{IJ}$.\footnote{Note that $E^I_i$ is invertible and $\mathcal{O}(r^0)$ so we deduce the term in brackets in \eqref{C} is $\mathcal{O}(r^{-N})$.} This is a second order differential equation in $r$ which means that there are two integration constants in $r$. The first corresponds to an $\mathcal{O}(r^0)$ term in the $C^I$ expansion, which must be zero by our fall-off condition $C^I = o(r^{-1})$. However, the second constant is genuine. For the analytic expansion we consider here, the $\mathcal{O}(r^{-3})$ term is undetermined, which means $C_1^I$ is unconstrained for the equations thus far considered, consistent with Ref.\ \cite{fakenews} . It is also consistent with the results of Ref.\ \cite{godazgar2020bms} for polyhomogeneous spacetimes. 
As discussed above, this corresponds to initial data and is in fact related to the angular momentum aspect  \cite{bondi,sachs}
\begin{equation}
L^I = -\frac{3}{2}C_1^I +\frac{3}{32}D^I H_{1JK}H_1^{JK} +\frac{3}{4} H_1^{IJ} D^K H_{1JK}.
\end{equation}

Next, we consider
\begin{align} \label{F} 
G_{01} &= -\frac{1}{r^2} e^{-2\beta} \partial_r (rF) +\frac{(h^{-1})^{ IJ}}{r^2} \big{(}(h^{-1})^{ KL} D_L h_{JK} D_I \beta -D_I D_J \beta -D_I \beta D_J \beta\big{)} \notag\\
&\hspace{20mm}+ \frac{e^{-2 \beta}}{2r^4} \partial_r (r^4 D_I C^I) -\frac{1}{4} e^{-4\beta} r^2 h_{IJ} \partial_r C^I \partial_r C^J \notag \\[2mm]
&\hspace{20mm}+\frac{1}{2r^2} \big{(} \Tr h +\tfrac{3}{4} (h^{-1})^{ IJ} D_I h_{KL} D_J (h^{-1})^{KL} +(h^{-1})^{ IJ} D_L ((h^{-1})^{KL} D_I h_{JK}) \notag\\[2mm]
&\hspace{20mm}+\frac{1}{2} (h^{-1})^{ KL} D_K (h^{-1})^{ IJ} D_I h_{JL}\big{)}.
\end{align}
Assuming $G_{01}=o(r^{-N})$ and the previous two equations, we obtain a first order differential equation for $F$ involving only $h_{IJ}$ and the aforementioned integration constant ($C_1^I$ for our expansion). This, in principle, allows one to determine $F$ up to a term constant in $r$ again. This constant is the coefficient of $r^{-1}$ in our $F$ expansion, which corresponds to $F_0$ in \eqref{metricexpansions}, related to the Bondi mass aspect $m=-\tfrac{1}{2}F_0$. This is again consistent with the polyhomogeneous case too. As discussed, this term corresponds to initial data as before. Importantly, from \eqref{F}, we can deduce that $F_0$ appears only once in the entire expansion for $F$, in the $r^{-1}$ coefficient. That is to say, for $n>0$, $F_n$ does not depend on $F_0$. This can be seen by noticing that \eqref{F} is invariant under $F\rightarrow F+ \delta F$, where $\delta F=\frac{\delta F_0(u,x^I)}{r}$, capturing the degree of freedom introduced by the integration constant.\footnote{In comparison, to leave $G_{0i}$ invariant in \eqref{C}, we would need ${C^I \rightarrow C^I +\delta C^I}$, where ${\delta C^I = \int dr  \tfrac{e^{2\beta} \delta C_1^J(u,x^I) (h^{-1})^I_J}{r^4}}$, contributing at all lower orders.}

So far, we have used the hypersurface equations to constrain the metric so that each term in the expansions on some initial hypersurface can only depend on $\{H_{(n\geq0)IJ}, F_0, C_1^I\}$. The remaining Einstein equations provide information on how these quantities evolve. We next consider $G_{ij}=o(r^{-(N-1)})$. This equation can be simplified for calculation purposes. First note that the Ricci scalar $R=\eta^{ab} R_{ab} = o(r^{-(N-1)})$, by our fall-off conditions, so that
\begin{align}
G_{ij} &= e_i^\mu e_j^\nu G_{\mu\nu} \ = \ \frac{1}{r^2} E_i^I E_j^J G_{IJ} \ =\ \frac{1}{r^2} E_i^I E_j^J( R_{IJ}-\tfrac{1}{2}R g_{IJ}) \notag \\[2mm]
&=  \frac{1}{r^2} E_i^I E_j^J R_{IJ} +o(r^{-(N-1)}).
\end{align}

\newpage
Therefore, the condition $G_{ij}=o(r^{-(N-1)})$ implies that $R_{IJ} = o(r^{-(N-3)})$, since $E_i^I = \mathcal{O}(r^0)$ and is invertible. Furthermore, we will only need to consider the traceless part $R_{\langle I J \rangle}$ as this contains all the new information at each order in $r$. We find that requiring that $R_{\langle I J \rangle} =  o(r^{-(N-3)})$ implies that
\begin{align}  \label{duh}
&re^{-2\beta} \partial_r ( r \partial_u h_{\langle IJ \rangle } ) - r^2 e^{-2\beta} (h^{-1})^{ KL} \partial_r h_{K \langle I} \partial_u h_{J \rangle L} \notag \\[2mm]
&=- \bigg{(} \frac{1}{4} D_{\langle I} (h^{-1})^{KL} D_{J \rangle} h_{KL} +D_{\langle I } h_{J\rangle K} D_L (h^{-1})^{ KL} -\frac{1}{2} D_K h_{\langle IJ\rangle } D_L (h^{-1})^{ KL} \notag\\
&- \frac{1}{2} (h^{-1})^{KL}  (h^{-1})^{ MN} D_L h_{\langle I | N} D_M h_{| J \rangle K} + \frac{1}{2} (h^{-1})^{ KL} (h^{-1})^{MN} D_M h_{\langle I | K} D_N h_{|J \rangle L} \notag\\
&+(h^{-1})^{KL} D_K D_{\langle I} h_{J\rangle L} -\frac{1}{2} (h^{-1})^{ KL} D_K D_L h_{\langle I J \rangle} \bigg{)}\notag\\
&-\bigg{(} (h^{-1})^{ KL} D_K \beta (2D_{\langle I} h_{J\rangle L} - D_L h_{\langle I  J\rangle } )-2D_{\langle I} D_{J \rangle} \beta -2D_{\langle I} \beta D_{J \rangle} \beta\bigg{)} \notag\\
&-r^2 e^{-2\beta}\bigg{(} -\frac{1}{2} e^{-2\beta}r^2 h_{K \langle I} h_{J \rangle L} \partial_r C^K \partial_r C^L + \frac{1}{r^2} \partial_r (r^2 h_{K\langle I} D_{J\rangle} C^K) \notag \\
&+\frac{1}{2} D_K h_{\langle I J \rangle} \partial_r C^K +\frac{1}{2} C^K \partial_r (rD_K h_{\langle I J \rangle}) +\frac{1}{2r^2} D_K C^K \partial_r (r^2 h_{\langle I J \rangle}) \notag\\
&-C^K (h^{-1})^{MN} D_K h_{M \langle I} \partial_r h_{J \rangle N} - (h^{-1})^{MN} D_N C^K h_{K \langle I} \partial_r h_{J \rangle M}\bigg{)} \notag\\
&+r^2 e^{-2\beta} \bigg{(} \frac{1}{2r^2} \partial_r (F \partial_r (r^2 h_{\langle I J \rangle})) - F\bigg{(}\frac{1}{r} \partial_r h_{\langle IJ \rangle} +\frac{1}{2} (h^{-1})^{ KL} \partial_r h_{K \langle I} \partial_r h_{J\rangle L} \bigg{)}\bigg{)} \notag\\
&+o(r^{-(N-3)}).
\end{align}
This is a first order equation in $r$ for $\partial_u h_{IJ}$. In principle, it can be solved to produce a first order evolution equation for $h_{IJ}$. The integration constant from the $r$ equation means that one cannot determine
\begin{equation}
\lim_{r\rightarrow \infty} (r \partial_u h_{\langle IJ\rangle }),
\end{equation}
which in our case is $\partial_u H_{1IJ}$, constituting free data. More typically denoted as $\partial_u C_{IJ}$, this is the Bondi news.

The final two equations we consider are the conservation equations $G_{11}=o(r^{-2})$ and $G_{1i}=o(r^{-3})$ providing the evolution equations for $F_0$ and $C_1^I$ respectively in terms of $H_{1IJ}$ and $F_0$
\begin{align}
\partial_u F_0 &= -\frac{1}{2} D_I D_J \partial_u {H_1}^{IJ} +\frac{1}{4} \partial_u H_{1IJ} \partial_u H_{1}^{IJ},  \label{duF0} \\[2mm]
\partial_u C_1^I &= \frac{1}{3} D^I F_0+\frac{1}{6}\Box D_I H_1^{IJ}-\frac{1}{6} D^I D^J D^K H_{1JK} +\frac{1}{8} H_{1JK} \partial_u D^I H_1^{JK} \notag\\
&+\frac{5}{8} \partial_u H_{1JK} D^I H_1^{JK} -\frac{2}{3} \partial_u H_{1JK} D^J H_1^{KI} -\frac{1}{6} D_J H_1^{IJ},  \label{duC1I}
\end{align}
where $\Box \equiv D_I D^I$ is the covariant Laplacian on the unit 2-sphere. \\

\subsection{Low order Einstein equations}
Now assuming our expansions in \eqref{metricexpansions}, the equations \eqref{beta}, \eqref{C}, \eqref{F} and \eqref{duh} can be solved order by order in $r^{-1}$. For sufficiently large $n$, it is possible to generalise the results needed in this paper, however, for smaller $n$, we need to compute the Einstein equations explicitly. Fortunately, this has been done previously in Ref.\ \cite{fakenews}, with the notation ${H_{1IJ} \equiv C_{IJ}}$, $H_{3IJ}\equiv D_{IJ}$ and $\bar{H}_{4IJ}\equiv E_{IJ}$. We will not reproduce them here and refer the reader to the equations of Ref.\ \cite{fakenews}.

\subsection{Linear Einstein equations}
In this section, we obtain general expressions for the linear terms appearing in Einstein's equations with the fall-off conditions \eqref{EinsteinFallOff} assumed. Firstly, considering the expression for $G_{00}$ in \eqref{beta}, one finds 
\begin{equation} \label{Blinear}
\beta_n \quad \text{is non-linear} \quad \text{for} \quad n \leq N-4.
\end{equation}

\noindent Evaluating \eqref{C} at $\mathcal{O}(r^{-n})$, one obtains
\begin{align} \label{Cn}
 \frac{1}{2} (n-1)(n-4) & C_{n-3}^I - \frac{1}{2} (n-2) D_J H^{IJ}_{n-2}+ \text{non-linear terms}=0 \quad \text{for} \quad 3\leq n\leq N \notag\\[2mm]
 \Rightarrow \quad & C_{n-3}^I= \frac{(n-2)}{(n-1)(n-4)} D_J H_{n-2}^{IJ}+ \text{non-linear terms}  \notag\\
& \hspace{40mm} \text{for} \quad n=3 \text{ and } 5\leq n\leq N.
\end{align}

\noindent Next, evaluating \eqref{F} at $\mathcal{O}(r^{-n})$, one obtains
\begin{align} \label{Flinear}
(n-3)&F_{n-3} -\frac{(n-5)}{2} D_I C_{n-3}^I +\frac{1}{2} D_I D_J H_{n-2}^{IJ}+ \text{non-linear terms}=0\notag\\
&\hspace{60mm}  \text{for} \quad 3\leq n\leq N \notag\\[2mm]
\Rightarrow \quad (n-3)&F_{n-3} = -\frac{(n-3)}{(n-1)(n-4)} D_I D_J H_{n-2}^{IJ}  + \text{non-linear terms}\notag\\
& \hspace{60mm} \text{for} \quad n=3 \text{ and } 5\leq n\leq N\notag\\[2mm]
\Rightarrow \quad  &F_{n-3}= -\frac{1}{(n-1)(n-4)} D_I D_J H_{n-2}^{IJ}+ \text{non-linear terms}\notag\\
&\hspace{60mm} \text{for} \quad 5\leq n\leq N,
\end{align}
where in the second line we used the $C_{n-3}^I$ equation \eqref{Cn} so the equation does not hold for $n=4$.

\noindent Finally, we look at the $\mathcal{O}(r^{-(n-2)})$ coefficient in \eqref{duh},
\begin{align} \label{Hlinear}
(n-2)\partial_u (H_{n-1})_{IJ} &=  (n-3)D_{\langle I} (C_{n-3})_{J\rangle} +\frac{1}{2} (n-2)(n-3) (H_{n-2})_{IJ}+ \text{non-linear terms} \notag\\ 
&\hspace{70mm} \text{for} \quad 4\leq n\leq N-1 \notag\\
\Rightarrow \partial_u (H_{n-1})_{IJ} &= -\frac{(n-3)}{(n-1)(n-4)} D_{\langle I |}D_K {H_{n-2 | J \rangle}}^K-\frac{1}{2}(n-3)(H_{n-2})_{ IJ} \notag\\ 
&\hspace{20mm}+ \text{non-linear terms} \quad \text{for} \quad 5\leq n\leq N-1,
\end{align}
where in the second line we have used the equation for $C_{n-3}^I$ in \eqref{Cn}.

\subsection{ $F_0$ terms}
As discussed in Section \ref{EinsteinStructure}, the $F_0$ term in \eqref{metricexpansions} appears only in the Einstein equations describing the evolution of the initial data $\{F_0, C_1^I, H_{(n\geq 3)IJ}\}$. In particular, we have
\begin{align} \label{F0terms}
\partial_u F_0 \big{|}_{F_0 \ \text{terms}} &= 0\notag\\[2mm]
\partial_u C_1^I \big{|}_{F_0\ \text{terms}} &= \frac{1}{3}D^I F_0\notag\\[2mm]
\partial_u H_{(n-1)IJ} \big{|}_{F_0 H_{(n-3)} \ \text{terms}} &= -\frac{(n-3)^2}{2(n-2)}F_0 H_{(n-3)IJ} \quad \text{for} \quad 3\leq n \leq N-1,
\end{align}
with the first two equations immediately following from \eqref{duF0} and \eqref{duC1I}. The third equation is obtained by considering the coefficient of $\mathcal{O}(r^{-(n-2)})$ in \eqref{duh}. The $F_0$ terms in the $H_n$ evolution equations
are non-linear and we will use these terms later to show that the higher order charges are not conserved for the non-linear theory.

\section{Charges} \label{Charges}
\subsection{BMS group}
We are considering asymptotically flat spacetimes, for which the asymptotic symmetry group is the BMS group \cite{bondi,Barnich:2009se}. This is the set of diffeomorphisms that preserve the form of the metric \eqref{metric} and \eqref{metricexpansions}. The group is parameterised by supertranslations and conformal Killing vectors on the 2-sphere. As in previous work \cite{fakenews,dualex,godazgar2020bms}, we will only focus on the supertranslation part of the BMS group as this is the novel feature. A general generator of such a diffeomorphism is then given by 
\begin{equation}
\xi= s\partial_u+\int dr \frac{e^{2\beta}}{r^2} (h^{-1})^{IJ} D_J s \hspace{1mm} \partial_I -\frac{r}{2} (D_I \xi^I - C^I D_I s ) \partial_r,
\end{equation}
where $s(x^I)$ is the supertranslation parameter. The variations of the metric functions can then be calculated using the Lie derivative. Below, we list the variations of functions for which there is no Einstein equation
\begin{align} \label{metvariations}
\delta F_0 &= s\partial_u F_0 -\tfrac{1}{2} \partial_u H_1^{IJ} D_I D_J s -D_I \partial_u H_1^{IJ} D_Js, \notag\\[2mm]
\delta  C_1^I &= s \partial_uC_1^I  +\tfrac{1}{16}\partial_u H_1^2 D^I s +F_0 D^I s -\tfrac{1}{4} H_1^{JK} D^I D_J D_K s -\tfrac{1}{2} H_1^{IJ} D_J \Box s \notag \\
&\hspace{10mm}+ \tfrac{1}{2} D^J H_1^{IK} D_J D_K s-\tfrac{3}{4} D^I H_1^{JK} D_J D_K s -\tfrac{1}{2}D_J H_1^{JK} D_K D^I s \notag\\
&\hspace{10mm}-\tfrac{1}{2}D^I D^J H_{1JK} D^K s+\tfrac{1}{2} D^J D_K H_1^{KI} D_J s-H_1^{IJ} D_J s,\notag \\[2mm]
\delta H_{1IJ} &=s\partial_u H_{1IJ} -2D_{\langle I} D_{J \rangle} s,\notag\\[2mm]
\delta H_{3IJ} &=s\partial_uH_{3IJ}-2C_{1\langle I} D_{J \rangle} s -\tfrac{1}{4} H_{1IJ} H_1^{KL} D_{K} D_{L} s-\tfrac{1}{8} H_1^2 D_{\langle I} D_{J \rangle} s \notag \\
&\hspace{10mm}+\tfrac{1}{8} D_{\langle I } H_1^2 D_{J\rangle} s +D_KH_1^{KL} H_{1L\langle I} D_{J \rangle} s, \notag\\[2mm]
\delta H_{nIJ} &= s\partial_u H_{nIJ} -\frac{(n-2)(n+1)}{n(n-3)} D_{\langle I} s D^K (H_{n-1})_{J \rangle K} - \frac{(n-2)}{n} D^K s D_{\langle I} (H_{n-1})_{J \rangle K} \notag\\
&\hspace{45mm}-\frac{(n-2)(n+1)}{n} D^K D_{\langle I} s (H_{n-1})_{J \rangle K}\notag\\
&\hspace{40mm}+ \text{non-linear terms} \quad \text{for} \quad 4\leq n\leq N-1,
\end{align}
where in the final expression, we have assumed the Einstein equation \eqref{Cn} which gives rise to our restriction on $n$. Here $H_1^2\equiv H_{1IJ} H_1^{IJ}$.

\subsection{Subleading charges}
It has recently been discovered that there are more charges associated with asymptotically flat spacetimes than previously thought. Charge variations can be derived from different contributions to the action in the tetrad formalism. For a discussion on this, we refer the reader to Ref.\ \cite{Godazgar:2020gqd}. Here, we will focus on the charges arising from the Palatini and Holst actions, responsible for the BMS charges and dual BMS charges respectively \cite{BarTro,dual0,dualex,Godazgar:2020kqd}
\begin{align} 
\slashed{\delta}\mathcal{Q}_\xi[ \delta g, g] &= {1\over{8\pi G}}  \int_S \star H[\xi,  g, \delta g] =  {1\over{8\pi G}}  \int_S d\Omega \hspace{1mm} r^2 e^{2\beta} H^{ur} [\xi,  g, \delta g]  \label{BBexpression}, \\[2mm]
\slashed{\delta}\mathcal{\widetilde{Q}}_\xi[ \delta g, g] &= {1\over{8\pi G}}  \int_S  \widetilde{H}[\xi,  g, \delta g] =  {1\over{8\pi G}}  \int_S d\Omega \hspace{1mm} \frac{\widetilde{H}_{\theta \phi} [\xi,  g, \delta g] }{\sin \theta} ,\label{DualCharge}
\end{align}
where we have used the form of the background metric of interest \eqref{metric} in the second equality in both equations. The slash on the variational symbol $\delta$ in \eqref{BBexpression} and \eqref{DualCharge} signifies the fact that the variation is not, in general, integrable. 
\\

\noindent The 2-forms $H$ and $\widetilde{H}$ are given by
\begin{align}
H= &\frac{1}{2} \Big\{ \xi_\nu g^{\rho \sigma} \nabla_\mu \delta g_{\rho \sigma} -\xi_\nu \nabla^\rho \delta g_{\mu \rho} +\xi^\rho  \nabla_\nu \delta g_{\rho \mu} \notag   \\
&\hspace{20mm}  + \frac{1}{2} g^{\rho \sigma} \delta g_{\rho \sigma} \nabla_\nu \xi_\mu + \frac{1}{2} \delta g_{\nu \rho} (\nabla_\mu \xi^\rho - \nabla^\rho \xi_\mu) \Big\}dx^\mu \wedge dx^\nu, \\[2mm]
\widetilde{H}= &  \frac{1}{4}\delta g_{JK} \big{(} \nabla_I \xi^K + \nabla^K \xi_I\big{)}dx^I \wedge dx^J \label{tildeH}.
\end{align}
Eq.\ \eqref{DualCharge} can be written covariantly on the 2-sphere as
\begin{equation}
\slashed{\delta}\mathcal{\widetilde{Q}}_\xi[ \delta g, g] = {1\over{16\pi G}}  \int_S d\Omega \hspace{1mm}\epsilon^{IJ} \widetilde{H}_{IJ} [\xi,  g, \delta g] ,
\end{equation}
where $\epsilon_{IJ}$ is the alternating tensor on the 2-dimensional subspace. We shall consider higher order charges, that is to say, we shall be calculating the above expressions as series in $r^{-1}$ and considering each coefficient in turn. We write
\begin{equation} \label{BMSexpansion}
 \ndelta \mathcal{Q}_\xi[\delta g, g]= \sum_{n=0}^{N-3}\frac{\ndelta \mathcal{Q}_n}{r^n} + o\big{(}r^{-(N-3)}\big{)},
\end{equation}
\begin{equation} \label{Dualexpansion}
 \ndelta \mathcal{\widetilde{Q}}_\xi[\delta g, g]= \sum_{n=0}^{N-3}\frac{\ndelta \mathcal{\widetilde{Q}}_n}{r^n} + o\big{(}r^{-(N-3)}\big{)}.
\end{equation}

\noindent In the following, it will be useful to define the twist of a symmetric tensor $X_{IJ}$ \cite{dual0,dualex}
\begin{equation} \label{Xtwist}
\widetilde{X}^{IJ} = X_K{}^{(I} \epsilon^{J)K}, \qquad \epsilon_{IJ} =  
\begin{pmatrix}                                                                                0 & 1 \\ -1 & 0                                                              \end{pmatrix} \sin \theta.
\end{equation}
If $X^{IJ}$ is trace-free, then $X_K{}^{[I} \epsilon^{J]K} = 0$, so we can drop the symmetrisation in the definition \eqref{Xtwist}. Note that the tilde on the quantity $\widetilde{H}$ above is not to be confused with the twist of $H$. The two quantities are a priori unrelated.

Evaluating the charges for our metric \eqref{metric} yields an integrable and non-integrable piece at each order. The non-integrable piece corresponds to flux, with the integrable piece corresponding to a charge that is conserved in the absence of flux \cite{WZ}. In general, the flux can be made to vanish under physically reasonable conditions on the metric. In Newman and Penrose's derivation of gravitational charges, the metric is allowed to remain general and the flux is made to vanish for a particular choice of spherical harmonic, which recently has been shown to correspond to a particular choice of supertranslation parameter \cite{fakenews}.

\section{Linear Newman-Penrose charges} \label{LinearNP}
In this section, we obtain expressions for the linear Newman-Penrose charges in this formalism and review their conservation properties. Writing the Weyl tensor as $C_{\mu \nu \rho \sigma}$ and assuming $G_{00}=o(r^{-N})$, the first Newman-Penrose Weyl scalar is defined as
\begin{align} \label{Psi0a}
\Psi_0 &= e_0^\mu e_2^\nu e_0^\rho e_2^\sigma C_{\mu \nu \rho \sigma} \notag\\
&=E_2^I E_2^J \bigg{(}\frac{1}{4} (h^{-1})^{KL} \partial_r h_{IK} \partial_r h_{JL}-\frac{1}{2r^2} \partial_r (r^2 \partial_r h_{IJ}) \notag\\
&\hspace{20mm} -\frac{1}{16r} \partial_r (r^2 h_{IJ} ) \partial_r h_{KL} \partial_r h^{-1 KL} \bigg{)} + o(r^{-N}).
\end{align}
Considering an expansion of $\Psi_0$ of the form ($N\geq5$)
\begin{equation}
\Psi_0 = \sum_{n=0}^{N-5} \frac{\Psi_0^n}{r^{n+5}} + o(r^{-N})
\end{equation}
and evaluating the coefficient of $r^{-(n+5)}$ in \eqref{Psi0a}, we find that for $0\leq n\leq N-5$,
\begin{equation} \label{Psi0}
\Psi_0^n = -\frac{1}{2} (n+2)(n+3) \hat{E}^I_2  \hat{E}^J_2 (H_{n+3})_{IJ}+ \text{non-linear terms}.
\end{equation}
Defining the differential operators $\eth$ and $\bar{\eth}$ acting on a scalar $\eta$ of spin $s$ as \cite{NP61,Goldberg:1966uu}
\begin{align}
\eth \eta &=-\frac{1}{2}(1+i)\sin^n \theta \bigg{(}\partial_\theta -\frac{1}{\sin \theta} \partial_\phi\bigg{)} \bigg{(}\frac{ \eta}{\sin^n \theta}\bigg{)}, \notag\\
\bar{\eth} \eta &=-\frac{1}{2}(1-i)\frac{1}{\sin^n \theta} \bigg{(}\partial_\theta +\frac{1}{\sin \theta} \partial_\phi\bigg{)} \big{(}\sin^n \theta \hspace{1mm} \eta\big{)}
\end{align}
and noting that $\Psi_0$ has spin 2, we construct a spin 0 quantity given by
\begin{equation} \label{Psi0d}
 \bar{\eth}^2 \Psi_0^n = -\frac{1}{4} (n+2)(n+3) D_I D_J (H_{n+3}^{IJ} - i \widetilde{H}_{n+3}^{IJ} ) + \text{non-linear terms}.
\end{equation}
The linear Newman-Penrose charges can be written as \cite{NP}
\begin{align}
\mathcal{G}^{n,0}_{m} &= \int_S d\Omega \hspace{1mm} _2 \bar{Y}_{n+2,m} \Psi_0^{n+1}  \quad \text{for} \quad n\geq 0,
\end{align}
where $_s Y_{\ell,m}$ are the spin weight $s$ spherical harmonics. We can use the definition of the spin-weighted spherical harmonics and \eqref{Psi0d} to rewrite the charges in our formalism as follows
\begin{align} 
\mathcal{G}^{n,0}_{m} &=  \sqrt{\frac{n!}{(n+4)!}}\int_S d\Omega \hspace{1mm} \bar{\eth}^2 \bar{Y}_{n+2,m} \Psi_0^{n+1}  \quad \text{for} \quad n\geq 0 \notag\\
&=\sqrt{\frac{n!}{(n+4)!}}\int_S d\Omega \hspace{1mm}  \bar{Y}_{n+2,m} \bar{\eth}^2 \Psi_0^{n+1}  \quad \text{for} \quad n\geq 0 \notag\\
&= -\frac{(n+3)(n+4)\sqrt{n!}}{4\sqrt{(n+4)!}}  \int_S d\Omega \hspace{1mm}  \bar{Y}_{n+2,m} D_I D_J (H_{n+4}^{IJ}-i \widetilde{H}_{n+4}^{IJ} ) \notag\\
&\hspace{50mm} \text{for} \quad 0\leq n \leq N-6, \label{NPCharges}
\end{align}
where the overall factor is unimportant.

\subsection{Conservation of linear Newman-Penrose charges} \label{ConsOfLin}
Newman and Penrose demonstrated that their charges are linearly conserved, that is to say their $u$ derivative vanishes without the need to impose constraints on the metric. We shall verify this result in the Bondi gauge. Consider 
\begin{equation} \label{Gn0}
\mathcal{G}^{n,0}_{m} = \frac{(n+3)(n+4)\sqrt{n!}}{4\sqrt{(n+4)!}} \big{(} \mathcal{Q}_{n+3,m} -i  \widetilde{\mathcal{Q}}_{n+3,m}\big{)}  \quad \text{for} \quad 0\leq n \leq N-6,
\end{equation}
where
\begin{gather} \label{Qn0}
\mathcal{Q}_{n,m} =-\int_S d\Omega\hspace{1mm} \bar{Y}_{n-1,m}D_I D_J H^{IJ}_{n+1} \quad \text{and} \quad
\widetilde{\mathcal{Q}}_{n,m} =-\int_S d\Omega\hspace{1mm} \bar{Y}_{n-1,m}D_I D_J \widetilde{H}^{IJ}_{n+1},
\end{gather}
for $3 \leq  n \leq N-3$, with both $\mathcal{Q}_{n,m}$ and $\widetilde{\mathcal{Q}}_{n,m} $ real charges. We will now show that each of these sets of charges are linearly conserved. Furthermore, we will replace $\bar{Y}_{n+2,m}$ with $s$, a generic function on the 2-sphere and show that the charges are conserved precisely when $s$ is a superposition of the appropriate spherical harmonics determined in the Newman-Penrose formalism. In the proceeding calculation, we will drop non-linear terms.  For $3\leq n \leq N-3$,
\begin{align}
\mathcal{Q}_{n}[s] &\equiv - \int_S d \Omega \hspace{1mm} D_I D_J s\hspace{1mm}  H_{n+1}^{IJ}  \label{Qns} \\ 
\Rightarrow \partial_u \mathcal{Q}_{n}[s] &=- \int_S d \Omega \hspace{1mm} D_I D_J s\hspace{1mm}  (\partial_u H_{n+1}^{IJ} ) \notag\\
&= \frac{n-1}{2(n+1)(n-2)} \int_S d \Omega \hspace{1mm} D_{\langle I} D_{J\rangle} s\hspace{1mm}  \big{(} 2 D^I D_K H_n^{JK} +(n+1)(n-2)H_n^{IJ} \big{)} \notag\\
&= \frac{n-1}{2(n+1)(n-2)} \int_S d \Omega \hspace{1mm} D_{\langle I} D_{J\rangle} s\hspace{1mm}  \big{(} \Box + (n^2-n-4) \big{)}H_n^{IJ} \notag\\
&= \frac{n-1}{2(n+1)(n-2)} \int_S d \Omega \hspace{1mm}  H_n^{IJ}  \big{(} \Box + (n^2-n-4) \big{)} D_{\langle I} D_{J\rangle} s,
\end{align}
where we have used \eqref{Hlinear} in going from the second to third line, a Schouten identity in going from the third to fourth (see App.\ B of Ref.\ \cite{godazgar2020bms}) and integration by parts in the final line. This expression is zero for arbitrary $H_n^{IJ}$ if and only if
\begin{equation}
T_{IJ} \equiv  \big{(} \Box + (n^2-n-4) \big{)} D_{\langle I} D_{J\rangle} s=0.
\end{equation}
This is a constraint on $s$. The condition $T_{IJ}=0$ is equivalent to the requirement that $\int_S d\Omega \hspace{1mm} |T_{IJ}|^2=0$. After integration by parts and repeated use of the Ricci identity, one can show that
\begin{equation}
\int_S d\Omega \hspace{1mm} |T_{IJ}|^2 = \frac{1}{2} \int_S d\Omega \hspace{1mm} s \Box (\Box+2) ( \Box + n(n-1))^2 s. 
\end{equation}
If we expand $s=\sum_{\ell,m} s_{\ell,m} Y_{\ell,m}$, in spherical harmonics, we can use the defining property $\Box Y_{\ell,m}= -\ell(\ell+1) Y_{\ell,m}$ and the orthogonality relation of the spherical harmonics to show
\begin{equation}
\int_S d\Omega \hspace{1mm} |T_{IJ}|^2 = \frac{1}{2} \sum_{\ell,m} |s_{\ell,m}|^2 (\ell- (n-1))^2(\ell-1) \ell (\ell +1)(\ell+2) (\ell+n)^2.
\end{equation}
This is zero precisely when $s$ is a linear combination of $\ell=0,1$ and $n-1$ modes. Note that if $s$ is an $\ell=0$ or $1$ spherical harmonic, then $D_{\langle I} D_{J \rangle} s =0$ and the charge is hence trivial, as can be seen from \eqref{Qns}. Hence, the contribution from such modes can be ignored by the linearity of \eqref{Qns} in $s$. We conclude that $s$ must be a superposition of $ Y_{n-1,m}$ modes for $m=0,\pm 1,..., \pm(n-1)$.\footnote{Note that $\bar{Y}_{\ell,m} =Y_{\ell,-m}$, so our conclusion is consistent with Newman and Penrose.}

We now repeat this calculation on the imaginary part of \eqref{Gn0}, again dropping non-linear terms. For $3\leq n \leq N-3$, 
\begin{align}
\mathcal{\widetilde{Q}}_{n}[s] &\equiv - \int_S d \Omega \hspace{1mm} D_I D_J s \hspace{1mm} \widetilde{H}_{n+1}^{IJ} 
\label{QnsDual} \\ 
&= \int_S d \Omega \hspace{1mm} {\epsilon_I}^K D_{\langle K} D_{J \rangle} s\hspace{1mm}  H_{n+1}^{IJ} \notag\\
&= \int_S d \Omega \hspace{1mm} {\epsilon_I}^K D_{ K} D_{J } s\hspace{1mm}  H_{n+1}^{IJ}  \notag\\
\Rightarrow \partial_u \mathcal{\widetilde{Q}}_{n}[s] &= \int_S d \Omega \hspace{1mm} {\epsilon_I}^K D_{ K} D_{J } s\hspace{1mm}  \partial_u (H_{n+1}^{IJ})  \notag\\
&= -\frac{n-1}{2(n+1)(n-2)} \int_S d \Omega \hspace{1mm}  {\epsilon_I}^K D_{ K} D_{J } s\hspace{1mm} \big{(} 2 D^{\langle I} D_K H_n^{J\rangle K} +(n+1)(n-2)H_n^{IJ} \big{)} \notag\\
&=- \frac{n-1}{2(n+1)(n-2)} \int_S d \Omega \hspace{1mm}  {\epsilon_I}^K D_{ K} D_{J } s \hspace{1mm}  \big{(} \Box + (n^2-n-4) \big{)}H_n^{IJ} \notag\\
&= -\frac{n-1}{2(n+1)(n-2)} \int_S d \Omega \hspace{1mm}  H_n^{IJ}  {\epsilon_I}^K \big{(} \Box + (n^2-n-4) \big{)}  D_{ K} D_{J } s \notag \\
&= \frac{n-1}{2(n+1)(n-2)} \int_S d \Omega \hspace{1mm}  \widetilde{H}_n^{JK} \big{(} \Box + (n^2-n-4) \big{)}  D_{ K} D_{J } s.
\end{align}
Now the same argument works as above since $\widetilde{H}_n^{IJ}$ is an arbitrary symmetric, traceless tensor. Hence, we deduce that \eqref{Gn0} are conserved and furthermore, they form a basis of all conserved charges that can be constructed in this manner.

\subsection{Newman-Penrose charges in the tetrad formalism}
It has been shown in Refs.\ \cite{fakenews,dualex} that the non-linearly conserved Newman-Penrose charges naturally arise as subleading charges in \eqref{BBexpression} and \eqref{DualCharge}, explaining their existence with regards to asymptotic symmetries. It is natural to consider this expansion further, with the expectation that additional Newman-Penrose charges arise. Indeed, at $\mathcal{O}(r^{-n})$ in the expansions \eqref{BMSexpansion} and \eqref{Dualexpansion}, for $n \leq N-3$, up to total derivatives, one finds\footnote{At this stage, neglecting the issue of how to pick the integrable piece, these expressions are valid in the full non-linear theory.}
\begin{align}
\slashed{\delta} \mathcal{Q}_n[s] &=  \int_S d \Omega \hspace{1mm} \delta \big{(}- s \hspace{1mm} (2F_n+nD_I C_n^I) \big{)}\quad + \quad \slashed{\delta} \mathcal{Q}^{(non-int)}_n[s],\\
\slashed{\delta} \mathcal{\widetilde{Q}}_n[s] &=  \int_S d \Omega \hspace{1mm} \delta \big{(}- D_I D_J s \hspace{1mm} \widetilde{H}_{n+1}^{IJ} \big{)}\quad + \quad \slashed{\delta} \mathcal{\widetilde{Q}}^{(non-int)}_n[s].
\end{align}
Appealing to the lower order Einstein equations in \cite{fakenews} and the linear equations \eqref{Cn} and \eqref{Flinear}, the first equation can be re-written, up to total derivatives, as
\begin{align}
\slashed{\delta} \mathcal{Q}_n[s] &=  \int_S d \Omega \hspace{1mm} \delta \big{(}- D_I D_J s \hspace{1mm} H_{n+1}^{IJ} \big{)}\quad + \quad \slashed{\delta} \mathcal{Q}^{(non-int)}_n[s],
\end{align}
for $2\leq n \leq N-3$. It should be noted that the separation into the integrable and non-integrable parts is somewhat arbitrary \cite{WZ}. At linear order, after using Einstein's equations and \eqref{metvariations}, the expressions for the metric variations in terms of $s$, one obtains
\begin{align}
 \slashed{\delta} \mathcal{Q}^{(non-int)}_n[s] &= -\int_S d \Omega \hspace{1mm} D_{\langle I} D_{J \rangle} s \bigg{(} \frac{n-1}{(n+1)(n-2)}s D^I D_K H_n^{JK}\notag\\
 &+\frac{(n-1)(n+2)}{n+1}H_n^{JK} D_K D^I s +\frac{1}{2} (n-1) s H_n^{IJ} \notag\\
 &+\frac{(n-1)(n+2)}{(n+1)(n-2)}D^I s D_K H_{n}^{JK} +\frac{n-1}{n+1}D_K s D^I H_n^{JK}\bigg{)}\notag\\
 &\hspace{50mm} + \text{non-linear terms},   \label{BMSnonint}\\
 \slashed{\delta} \mathcal{\widetilde{Q}}^{(non-int)}_n[s] &= -\int_S d \Omega \hspace{1mm} D_{\langle I} D_{J \rangle} s \bigg{(} \frac{n-1}{(n+1)(n-2)}s D^I D_K \widetilde{H}_n^{JK}\notag\\
 &+\frac{(n-1)(n+2)}{n+1}\widetilde{H}_n^{JK} D_K D^I s +\frac{1}{2} (n-1) s \widetilde{H}_n^{IJ} \notag\\
 &+\frac{(n-1)(n+2)}{(n+1)(n-2)}D^I s D_K \widetilde{H}_{n}^{JK} +\frac{n-1}{n+1}D_K s D^I \widetilde{H}_n^{JK}\bigg{)}\notag\\
 &\hspace{50mm} + \text{non-linear terms}, \label{Dualnonint}
\end{align}
for $3\leq n \leq N-3$. It is now a matter of determining for each $n$ whether these expressions can be made to vanish for particular choices of $s$. This, however, only proves that the integrable charge is invariant under the action of the specific supertranslation used to define that particular charge. It does not prove full ST invariance. In order to do this, we need to consider the action of a general ST, generated by $s_2$ say, on the charges defined using $s_1$.\footnote{Note that full ST invariance of the non-linear $\mathcal{O}(r^{-3})$ Newman-Penrose charges is demonstrated in this formalism in App.\ A of \cite{fakenews}.}
\newpage

\noindent The variation is closely related to \eqref{BMSnonint} and \eqref{Dualnonint} and is given by
\begin{align}
 \slashed{\delta}_{s_2} \mathcal{Q}^{(non-int)}_n[s_1] &= \int_S d \Omega \hspace{1mm} D_{\langle I} D_{J \rangle} s_1 \bigg{(} \frac{n-1}{(n+1)(n-2)}s_2 D^I D_K H_n^{JK}\notag\\
 &+\frac{(n-1)(n+2)}{n+1}H_n^{JK} D_K D^I s_2 +\frac{1}{2} (n-1) s_2 H_n^{IJ} \notag\\
 &+\frac{(n-1)(n+2)}{(n+1)(n-2)}D^I s_2 D_K H_{n}^{JK} +\frac{n-1}{n+1}D_K s_2 D^I H_n^{JK}\bigg{)}\notag\\
 & + \text{non-linear terms}, \\
 \slashed{\delta}_{s_2} \mathcal{\widetilde{Q}}^{(non-int)}_n[s_1] &= \int_S d \Omega \hspace{1mm} D_{\langle I} D_{J \rangle} s_1 \bigg{(} \frac{n-1}{(n+1)(n-2)}s_2 D^I D_K \widetilde{H}_n^{JK}\notag\\
 &+\frac{(n-1)(n+2)}{n+1}\widetilde{H}_n^{JK} D_K D^I s_2 +\frac{1}{2} (n-1) s_2 \widetilde{H}_n^{IJ} \notag\\
 &+\frac{(n-1)(n+2)}{(n+1)(n-2)}D^I s_2 D_K \widetilde{H}_{n}^{JK} +\frac{n-1}{n+1}D_K s_2 D^I \widetilde{H}_n^{JK}\bigg{)}\notag\\
 & + \text{non-linear terms}.
 \end{align}
For each $n$, to have an ST invariant charge, we need some fixed $s_1$ to exist, where the expression on the right vanishes for all $s_2$. This further implies the RHS of \eqref{BMSnonint} and \eqref{Dualnonint} vanish for this $s_1$ simply by noting the case $s_2=-s_1$ will have already been proven.

We start by noting the case in which $s_2= const$ corresponds to $\xi \sim \partial_u$ which is the matter considered in the previous subsection. We hence deduce that $s_1$ must be a superposition of $\ell=n-1$ spherical harmonics for both sets of charges. We shall hence write $s_1 \equiv S_{n-1}$ where $\Box S_{n-1} = -n(n-1)S_{n-1}$. For convenience, we we shall drop the subscript on $s_2$. Now, up to total derivatives, the expressions above can be re-written
\begin{align}
 \slashed{\delta}_{s} \mathcal{Q}^{(non-int)}_n[S_{n-1}] &= \frac{n-1}{2(n+1)(n-2)} \int_S d \Omega \hspace{1mm} H_n^{IJ} \bigg{(}-2(n+1) D^K s D_I D_{\langle J} D_{K \rangle} S_{n-1} \notag\\
 &+2(n-3)(n+1)D_I D^K s D_{\langle J} D_{K \rangle} S_{n-1}-2(n-3)D_I s D^K D_{\langle J} D_{K \rangle} S_{n-1} \notag\\
 &+2sD_I D^K D_{\langle J} D_{K \rangle} S_{n-1}+(n+1)(n-2) s D_I D_J S_{n-1}\bigg{)}  \notag\\
 & + \text{non-linear terms},\\
  \slashed{\delta}_{s} \mathcal{\widetilde{Q}}^{(non-int)}_n[S_{n-1}] &= \frac{n-1}{2(n+1)(n-2)} \int_S d \Omega \hspace{1mm} \widetilde{H}_n^{IJ} \bigg{(}-2(n+1) D^K s D_I D_{\langle J} D_{K \rangle} S_{n-1} \notag\\
 &+2(n-3)(n+1)D_I D^K s D_{\langle J} D_{K \rangle} S_{n-1}-2(n-3)D_I s D^K D_{\langle J} D_{K \rangle} S_{n-1} \notag\\
 &+2sD_I D^K D_{\langle J} D_{K \rangle} S_{n-1}+(n+1)(n-2) s D_I D_J S_{n-1}\bigg{)}  \notag\\
  &+ \text{non-linear terms}.
 \end{align}

 \newpage
\noindent We want the linear part of these expressions to vanish for all $s$ for arbitrary symmetric traceless $H_n^{IJ}$ and $\widetilde{H}_n^{IJ}$. In both cases, this is true iff $X_{\langle I J \rangle}[s;S_{n-1}]=0$ where\footnote{Recall that $n\geq 3$ so the overall factor is non-zero.}
\begin{align}
X_{ I J }[s;S_{n-1}]  &= -2(n+1) D^K s D_I D_{\langle J} D_{K \rangle} S_{n-1} \notag\\
 &+2(n-3)(n+1)D_I D^K s D_{\langle J} D_{K \rangle} S_{n-1}-2(n-3)D_I s D^K D_{\langle J} D_{K \rangle} S_{n-1} \notag\\
 &+2sD_I D^K D_{\langle J} D_{K \rangle} S_{n-1}+(n+1)(n-2) s D_I D_J S_{n-1}. \label{X1}
 \end{align}
This expression can be simplified using the following identity
\begin{align}
&D^K D_{\langle J} D_{K \rangle} S_{n-1} = -\frac{1}{2} (n-2)(n+1) D_J S_{n-1}. \label{Ricci1}
 \end{align}
This follows by using the Ricci identity and then the property that $S_{n-1}$ is an $\ell=n-1$ spherical harmonic. One can then deduce that the final line of \eqref{X1} is zero. Furthermore, the third term can be simplified leaving us with
\begin{align}
X_{ I J }[s;S_{n-1}]  &= (n+1)\big{(}-2 D^K s D_I D_{\langle J} D_{K \rangle} S_{n-1} \notag\\
 &+2(n-3)D_I D^K s D_{\langle J} D_{K \rangle} S_{n-1}+(n-3)(n-2)D_I s D_J S_{n-1} \big{)}. \label{X2}
 \end{align}
For $n=3$, the final two terms clearly vanish for all $s$. Things are not immediately clear for the first term, but it can be shown that the trace-free symmetric part of the differential operator acting on $S_{n-1}$ gives zero precisely when acting upon a composition of $\ell=0,1$ and $2$ spherical harmonics\footnote{This is an adaptation of the arguments in App.\ C of \cite{fakenews} and App.\ C of \cite{godazgar2020bms} with ${T_{IJK}=(D_{\langle I} D_{J\rangle} D_K -\tfrac{1}{2}\gamma_{K \langle I} D_{J\rangle} \Box)s=0}$.}, hence $(D_{\langle I} D_{J\rangle} D_K -\tfrac{1}{2}\gamma_{K \langle I} D_{J\rangle} \Box)S_{2}=0$. Therefore, we have a non-trivial, supertranslation invariant charge at the linear level. This charge is shown to be non-linearly supertranslation invariant in App.\ A of \cite{fakenews}.

We now focus on $n>3$. For $X_{\langle I J \rangle}[s;S_{n-1}]=0$, it is necessary that for any constant symmetric, trace-free tensor $\alpha_{IJ}$,
\begin{align}
\int_S d \Omega \hspace{1mm} \alpha^{IJ} X_{IJ}=0.
 \end{align}
Using the expression for $X_{IJ}$ in \eqref{X2} and integrating by parts, this can be written as
\begin{align}
\int_S d \Omega \hspace{1mm} \alpha^{IJ} X_{IJ}& = (n+1)(n-2) \int_S d \Omega \hspace{1mm} \alpha^{IJ}  s \big{(} 2D^K D_I D_{\langle J} D_{K \rangle} S_{n-1} -(n-3) D_I D_J S_{n-1}\big{)} \notag\\
&= (n+1)(n-2) \int_S d \Omega \hspace{1mm} \alpha^{IJ}  s \big{(} D_I D_J \Box S_{n-1} -(n-9)D_I D_J S_{n-1} \big{)} \notag \\
&= (n+3)(n+1)(n-2)(n-3) \int_S d \Omega \hspace{1mm} \alpha^{IJ}  s \hspace{1mm} D_I D_J S_{n-1}, \end{align}
where in going from the first to the second line, we have used the Ricci identity and in going from the second to the third line, we have used $\Box S_{n-1} = - n(n-1) S_{n-1}$. For $n>3$, the overall factor does not vanish, so for the expression to be zero for all $s$ and all $\alpha_{IJ}$, we need $D_{\langle I} D_{J \rangle} S_{n-1}= 0$. This is true iff $n=1$ or $2$. Hence, at the linear level, the only ST invariant Newman-Penrose charges are the 5 complex charges $\mathcal{G}^{0,0}_{m}$, with $m=0,\pm 1, \pm 2$. These are the charges that are conserved in the full non-linear theory. For completeness, a demonstration of the ST invariance of the BMS and dual BMS charges is provided in Appendix \ref{app:STBMS}.

\section{Non-linear conservation of Newman-Penrose charges} \label{NLNP}
The full non-linear account of the tetrad formalism for the Palatini and Holst actions has been done in Refs.\ \cite{fakenews,dualex} for $n\leq 3$. In particular, at $\mathcal{O}(r^{0})$, the BMS and dual charges are discovered and at $\mathcal{O}(r^{-3})$, the Newman-Penrose charges are found. Newman and Penrose considered $\mathcal{O}(r^{-4})$ and showed that their argument used at the previous order for conservation fails here. In this section, we will present a demonstration of what happens beyond this order by considering specific terms in the non-integrable piece that cannot be made to vanish using the Einstein equations.

We will now demonstrate that the Newman-Penrose charges are not non-linearly conserved for $n>3$. In principle, we could begin this discussion as we did in Section \ref{ConsOfLin} by setting $\bar{Y}_{n+2,m} $ to be general $s$ again; however, this is not necessary. We have learned that in order for the linear terms, involving $H_{n+1}^{IJ}$, in $\partial_u \mathcal{Q}_{n}$ to vanish, it is necessary for $s$ to be a linear combination of one of the spherical harmonics found by Newman and Penrose. Seeing as there is no Einstein equation relating $H_{n+1}^{IJ}$ to anything else when it does not appear as a $u$ derivative, there is no chance of the additional non-linear terms possibly cancelling these contributions out. Therefore, for $3<n\leq N-3$, the charges must be of the form
\begin{align}
\mathcal{Q}_{n}[S_{n-1}] \equiv - \int_S d \Omega \hspace{1mm} D_I D_J S_{n-1} \hspace{1mm}  H_{n+1}^{IJ}, \\
\mathcal{\widetilde{Q}}_{n}[S_{n-1}] \equiv - \int_S d \Omega \hspace{1mm} D_I D_J S_{n-1} \hspace{1mm}  \widetilde{H}_{n+1}^{IJ},
\end{align}
where $S_{n-1}$ is a linear combination of $\ell=n-1$ spherical harmonics.

We shall demonstrate that these charges are not conserved by considering terms in $\partial_u \mathcal{Q}_{n}$ and $\partial_u \mathcal{\widetilde{Q}}_{n}$ that contain both $F_0$ and $H_{n-1}^{IJ}$ after the use of Einstein's equations. Seeing as there is no Einstein equation for either of these terms (when they do not appear as $u$ derivatives), the $\sim F_0 \hspace{1mm} H_{n-1}^{IJ}$ terms need to vanish separately in order for the charge to be conserved. Using  \eqref{duh} and \eqref{F0terms}, we find that
\begin{align} \label{F0termsinCharges1}
\partial_u \mathcal{Q}_{n} \big{|}_{F_0 H_{n-1}^{IJ} \text{terms}} &= -\frac{(n-1)^2}{2n}  \int_S d \Omega \hspace{1mm} F_0  \hspace{1mm}  H_{n-1}^{IJ} D_I D_J S_{n-1}, \\
\partial_u \mathcal{\widetilde{Q}}_{n} \big{|}_{F_0 H_{n-1}^{IJ} \text{terms}} &= -\frac{(n-1)^2}{2n}  \int_S d \Omega \hspace{1mm} F_0  \hspace{1mm}  \widetilde{H}_{n-1}^{IJ} D_I D_J S_{n-1},  \label{F0termsinCharges2}
\end{align}
which is actually valid for $1\leq n \leq N-3$ if one extends the definition of charges to $n=1,2$ using the integrable pieces found in Refs.\ \cite{fakenews,dualex}. For $n=1$, these terms vanish so there are no $F_0$ terms at all in $\mathcal{Q}_{1}$ and  $\mathcal{\widetilde{Q}}_{1}$. For $n=2$, we get the non-zero term $-\frac{1}{4}F_0 C^{IJ} D_I D_J S_1$\footnote{Recall that the more standard notation is $H_1^{IJ} \equiv C^{IJ}$.} in $\mathcal{Q}_{2}$ and $-\frac{1}{4}F_0 \widetilde{C}^{IJ} D_I D_J S_1$ in $\mathcal{\widetilde{Q}}_{2}$. Both terms vanish precisely since $S_1$ is a superposition of $\ell=0,1$ spherical harmonics, which in both cases make the charge trivial. These results are all consistent with those found in Refs.\ \cite{fakenews,dualex}. It is $n=3$ where the magic happens. Due to the peeling condition, $H_{2IJ}=0$ so the RHS of \eqref{F0termsinCharges1} and \eqref{F0termsinCharges2} vanish and there are no such terms in the charge derivatives to potentially destroy conservation. In order to prove these charges really are conserved, one has to consider all terms appearing in the charge derivatives. Fortunately, this has been done in \cite{NP} (see also \cite{fakenews,dualex}). For $n>3$, for general $H_{nIJ}$ and $F_0$, both \eqref{F0termsinCharges1} and \eqref{F0termsinCharges2} are non-zero and hence the charges cannot possibly be conserved. The introduction of this non-linear term alone ruins the conservation of the higher order Newman-Penrose charges.

It is unusual that the peeling condition is what protects the 10 non-linearly conserved Newman-Penrose charges given it has been shown that relaxing the peeling condition still gives rise to conserved charges at this order in $r$. For a more detailed discussion on this in this formalism, we refer the reader to Ref.\  \cite{godazgar2020bms}.

This analysis justifies Newman and Penrose's claim that there are no further charges of this form. Seeing as the charges are not conserved in $u$ for $n>3$, it follows immediately that the charges are not supertranslation invariant.\footnote{Recall that the supertranslation with parameter $s=const$ is in fact $\xi \propto \partial_u$. The charge is invariant under this paramter iff it is conserved in $u$.}

\section{Discussion} \label{sec:dis}
In this paper, we have further investigated whether or not there are further Newman-Penrose charges beyond the existing 10. This can be seen as a step towards the conclusion of the investigation carried out in Refs.\  \cite{fakenews} and \ \cite{dualex}, extending the arguments to all orders in $r^{-1}$ within the realm of what is mathematically viable. We verified that the linear Newman-Penrose charges are indeed conserved with respect to $u$ in this formalism, as ought to have been the case given Newman and Penrose's findings. However, the quantities are not charges in the sense of the tetrad formalism and in particular, do not exhibit supertranslation invariance. It has already been shown in Ref.\ \cite{fakenews} that the 10 non-linear Newman-Penrose charges are fully supertranslation invariant, as was also shown by Newman and Penrose.

Although hope is lost in showing that the linear Newman-Penrose charges have a place in this formalism, there remains several other possibilities that one can argue. It is possible that the non-integrable pieces can be made to vanish by instead restricting the metric in the same fashion as one does with the BMS and leading order dual BMS charges. These charges exist for any $s$ and are conserved in the absence of Bondi news, i.e.\ $\partial_u C_{IJ}=0$. It is possible that imposing similar constraints on the metric may result in the non-integrable piece vanishing at some other order of $r^{-1}$, however, seeing as there is no Einstein equation relating $H_{nIJ}$ to other terms (except when it appears as a $u$ derivative), the arguments in this paper regarding the linear terms in the non-integrable piece still hold and we would require conditions on the metric such as
\begin{equation}
D_I D_J(\Box+(n^2-n-4))H_n^{IJ}=0 \quad \text{and} \quad D_I D_J(\Box+(n^2-n-4))\widetilde{H} _n^{IJ}=0,
\end{equation}
for which it is difficult to find physical meaning, not to mention that these would be a subset of many further requirements when one considers the entire non-integrable piece.

Another possibility is that the separation into the integrable and non-integrable piece may be adjusted. Perhaps by moving terms from the non-integrable piece into the integrable piece, it is possible to remove the $F_0$ terms in such a way that the non-integrable piece can be made to vanish for some $s$. Given the rapidly increasing complexity of the equations at each order, it is difficult to believe that this will work, but it cannot be ruled out without having a clearer understanding of how one chooses the integrable piece in this formalism, a problem that is still to be addressed at subleading orders.

In summary, we have shown that although the infinite tower of non-linearly conserved Newman-Penrose charges do appear very naturally in the tetrad formalism, they are not charges in the traditional sense. That being said, one can argue that the charges are still of mathematical significance. In particular in \cite{Mao:2020vgh}, the infinite tower is realised with a different origin, motivated by gravitational memory effects \cite{zel1974radiation,braginskii1985kinematic,braginsky1987gravitational,christodoulou1991nonlinear}.

Since their discovery, the charges' conservation properties have been shown to hold true in even broader classes of spacetimes with slower fall-off conditions on the Weyl scalar \cite{valiente1999logarithmic,godazgar2020bms}. The Newman-Penrose charges are a prominent feature of asymptotically flat spacetimes and this study further demonstrates their significance through their individuality.

\section*{Acknowledgements}
I would like to thank Mahdi Godazgar for discussions that initiated this study and his support throughout this project. I am supported by a Royal Society Enhancement Award.

\appendix

\section{Supertranslation invariance of BMS and dual charges}  \label{app:STBMS}
We demonstrate the supertranslation invariance of the BMS and dual BMS charges, which in turn implies their conservation in $u$. We will write $H_{1IJ}\equiv C_{IJ}$ throughout. There is a pair of charges for each spherical harmonic parameterised by $\ell\geq 0$ and $m=0,\pm1,..,\pm\ell$,
\begin{align}
\mathcal{Q}_{\ell,m} &= -2\int_S d\Omega \hspace{1mm} Y_{\ell,m} \hspace{1mm} F_0, \\
\widetilde{\mathcal{Q}}_{\ell,m} &=  -\int_S d\Omega  \hspace{1mm} Y_{\ell,m} \hspace{1mm} D_I D_J \widetilde{C}^{IJ}.
\end{align}
From the expressions for the charge variations \eqref{metvariations} and Einstein equations \eqref{Blinear}, \eqref{Cn}, \eqref{Flinear} and \eqref{Hlinear}, we deduce that
\begin{align}
\delta \widetilde{C}_{IJ} &= s\partial_u \widetilde{C}_{IJ} -{\epsilon_{I}} ^K D_{\langle K} D_{J \rangle} s, \notag \\
\delta F_0 &= -\frac{1}{2} D_I D_J (s \partial_u C^{IJ} ) +\frac{1}{4}s \partial_u C_{IJ} \partial_u C^{IJ},
\end{align}
which make the calculations straightforward. The variation of the charges with respect to a supertranslation parameter $s$ are given by
\begin{align}
\delta_s \mathcal{Q}_{\ell,m} &= \int_S d\Omega \hspace{1mm} Y_{\ell,m} \bigg{(} D_I D_J (s \partial_u C^{IJ}) -\frac{1}{2} s \partial_u C_{IJ} \partial_u C^{IJ}\bigg{)}, \label{BMSvar}\\
\delta_s \widetilde{\mathcal{Q}}_{\ell,m} &=  \int_S d\Omega  \hspace{1mm} Y_{\ell,m} \hspace{1mm} D_I D_J \bigg{(}-s \partial_u \widetilde{C}^{IJ} + {\epsilon^I}_K D^{\langle K} D^{J \rangle} s \bigg{)} . \label{dualvar}
\end{align}
In \eqref{dualvar}, the second term can be shown to vanish by the Ricci identity. The first term vanishes if $\partial_u C_{IJ}=0$ or if $\ell=0$ or $1$, after integration by parts. However, in the latter case, the charge is trivial. The variations of each set of non-trivial charges are zero iff $\partial_u C_{IJ}=0$. Hence, the non-trivial charges are supertranslation invariant precisely when the Bondi news tensor vanishes, i.e.\ in the absence of flux at null infinity.

\newpage

\bibliographystyle{utphys}
\bibliography{NP}

\end{document}